\documentclass[titlepage,12pt]{article}
\usepackage{amssymb}
\usepackage{amsfonts}
\usepackage{amsmath}

\begin{document}

\title{An unsuitable use of spin and pseudospin symmetries with a
pseudoscalar Cornell potential}
\author{L. B. Castro\thanks{%
luis.castro@pgfsc.ufsc.br } \\
Departamento de F\'{\i}sica, CFM\\
Universidade Federal de Santa Catarina,\\
88040-900, CP. 476, Florian\'{o}polis - SC, Brazil\\\\
and\\\\
A. S. de Castro\thanks{%
castro@pesquisador.cnpq.br}\\
Departamento de F\'{\i}sica e Qu\'{\i}mica, Campus de Guaratin\-gue\-t\'{a}\\
Universidade Estadual Paulista,\\
12516-410, Guaratinguet\'{a} - SP, Brazil}
\date{}
\maketitle

\begin{abstract}
It is shown that a recent publication (2013 \textit{Chin. Phys. B} \textbf{22} 090301) uses
the concepts of spin and pseudospin symmetries as mere rhetoric to decorate
the pseudoscalar potential. It is also pointed out that a more complete
analysis of the bound states of fermions in a pseudoscalar Cornell potential
had already been published elsewhere.

\bigskip


\bigskip

\end{abstract}

Back in 2004 de Castro \cite{asc1} found the analytic solutions of the Dirac
equation for a pseudoscalar linear plus Coulomb-like potential%
\begin{equation}
mc\omega x+\frac{\hbar cg}{x}
\end{equation}%
where $\omega $ and $g$ are real parameters. The author of Ref. \cite{asc1}
approached the problem under a Sturm-Liouville perspective for a large range
of parameters but the possibility of isolated solutions from the
Sturm-Liouville perspective was neglected \cite{asc2}. In a recent article
published in Chinese Physics B, Hamzavi and Rajabi \cite{ham} alleged to
have found the exact solutions of the Dirac equation with a pseudoscalar
Cornell potential%
\begin{equation}
-\frac{a}{x}+bx  \label{cor}
\end{equation}%
under the spin and pseudospin symmetry limits. The purpose of this work is
to clarify that Hamzavi and Rajabi failed to perceive the possibility of
isolated solutions, as did de Castro in Ref. \cite{asc1}. More than this,
Hamzavi and Rajabi incurred in a clear lack of understanding about spin and
pseudospin symmetries and truly presented the solutions for a pure
pseudoscalar coupling for a more restrict range of parameters than in Ref.
\cite{asc1}.

Spin and pseudospin symmetries are SU(2) symmetries of a Dirac equation with
vector and scalar potentials realized when the difference between the
potentials, or their sum, is a constant \cite{bel}. The upper (lower)
component of the Dirac spinor satisfies a Schr\"{o}dinger-like equation in
the case of spin (pseudospin) symmetry and the resulting spectrum is
independent of the orientation of the spin (see, e.g. \cite{gin}).
Furthermore, the spin-orbit and the Darwin terms disappear from the
second-order equations for the upper and lower components of the Dirac
spinor under the spin and pseudospin conditions in such a way that spin-1/2
and spin-0 particles have equivalent spectra \cite{ped}. If the difference
between the vector and scalar potentials, or their sum, is not a constant
there is no spin, or pseudospin, symmetry. Spin and pseudospin symmetries
are also broken if the potential is contaminated by other sorts of couplings
(pseudoscalar or tensor couplings, for example) \cite{ron}. In fact, there
has been a continuous interest for solving the Dirac equations in the
four-dimensional space-time as well as in lower dimensions by assuming that
the vector potential has the same magnitude as the scalar potential \cite%
{asc1}-\cite{aut46}. In 1+1 dimensions it is usual to say that the system has
spin or pseudospin symmetries if the potentials are limited to vector and
scalar Lorentz structures with the same magnitude because, despite the
absence of spin effects, many attributes of the real spin or pseudospin
symmetries in 3+1 dimensions are preserved. In the presence of
time-independent interactions the most general 1+1 dimensional
time-independent Dirac equation for a fermion of rest mass $m$ and momentum $%
p$ reads%
\begin{equation}
\left( \sigma _{1}\hat{p}+\sigma _{3}m+\frac{I+\sigma _{3}}{2}V_{\Sigma }+%
\frac{I-\sigma _{3}}{2}V_{\Delta }+\sigma _{2}V_{p}\right) \psi =E\psi
\label{dirac1}
\end{equation}%
where $E$ is the energy of the fermion, $V_{\Sigma }=V_{v}+V_{s}$ and $%
V_{\Delta }=V_{v}-V_{s}$. The subscripts for the terms of potential denote
their properties under a Lorentz transformation: $v$ for the time component
of the two-vector potential, $s$ for the scalar potential and $p$ for the
pseudoscalar potential. $\sigma _{i}$ are the Pauli matrices and $I$ is the 2%
$\times $2 unit matrix. Hamzavi and Rajabi \cite{ham} in the subsection
\textit{Spin symmetry limit }(subsection 3.1) considered an equation for the
upper component of the Dirac spinor with $\Sigma =0,$ $\Delta =C_{s}=0$, $%
E\neq -m$, and in the subsection \textit{Pseudospin symmetry limit }%
(subsection 3.2) considered an equation for the lower component of the Dirac
spinor with $\Delta =0$, $\Sigma =C_{ps}=0$, $E\neq +m$. Therefore, in both
circumstances they took into account $V_{v}=V_{s}=0$. Indeed, in simple
words, they in fact considered a pure pseudoscalar potential in such a way
that for $E\neq \mp m$
\begin{equation}
\psi _{\mp }=-\frac{i}{E\pm m}\left( \frac{d\psi _{\pm }}{dx}\mp V_{p}\psi
_{\pm }\right)
\end{equation}%
and%
\begin{equation}
-\frac{d^{2}\psi _{\pm }}{dx^{2}}+\left( V_{p}^{2}\pm \frac{dV_{p}}{dx}%
\right) \psi _{\pm }=\left( E^{2}-m^{2}\right) \psi _{\pm }
\end{equation}%
On the other hand, for $E=\mp m$ one has%
\begin{eqnarray}
\psi _{\pm } &=&N_{\pm }e^{_{\pm }v\left( x\right) }  \notag \\
&&  \label{q2} \\
\psi _{\mp } &=&\left[ N_{\mp }-iN_{\pm }I_{\pm }\left( x\right) \right]
e^{_{\mp }v\left( x\right) }  \notag
\end{eqnarray}%
where $N_{+}$ and $N_{-}$ are normalization constants, and%
\begin{eqnarray}
v\left( x\right)  &=&\int\nolimits^{x}dy\,V_{p}\left( y\right)   \notag \\
&&  \label{II} \\
I_{\pm }(x) &=&\pm 2m\int\nolimits^{x}dy\,e^{\pm 2v\left( y\right) }  \notag
\end{eqnarray}

In summary manner, the solutions of the Sturm-Liouville problem with a
pseudoscalar linear plus Coulomb-like potential (or Cornell potential if one
likes) had already been extensively discussed in Ref. \cite{asc1}, and the
isolated solutions in Ref. \cite{asc2}. What is more, the rhetoric is great
but there is no spin or pseudospin symmetries in the system approached in
Ref. \cite{ham} because of the complete absence of vector and scalar
couplings.


\bigskip

\begin{thebibliography}{9}
\bibitem{asc1} de Castro A S 2004 \textit{Ann. Phys. (N.Y.)} \textbf{311} 170

\bibitem{asc2} Castro L B and de Castro A S 2013 \textit{Ann. Phys. (N.Y.)} \textbf{338} 278

\bibitem{ham} Hamzavi M and Rajabi A A 2013 \textit{Chin. Phys. B} \textbf{22} 090301

\bibitem{bel} Bell J S and Ruegg H 1975 \textit{Nucl. Phys. B} \textbf{98} 151

\bibitem{gin} Ginocchio J N 2005 \textit{Phys. Rep. } \textbf{414} 165

\bibitem{ped} Alberto P, de Castro A S and Malheiro M 2007 \textit{Phys. Rev. C} \textbf{75} 047303

\bibitem{ron} Lisboa R, Malheiro M, de Castro A S, Alberto P and Fiolhais M 2004 \textit{Int. J. Mod. Phys. D} \textbf{13} 1447

\bibitem{aut1} Alberto P, Lisboa R, Malheiro M and de Castro A S 2005 \textit{Phys. Rev. C} \textbf{71} 034313

\bibitem{gum} Gumbs G and Kiang D 1986 \textit{Am. J. Phys.} \textbf{54} 462

\bibitem{aut2} Dom\'{\i}nguez-Adame F 1990 \textit{Am. J. Phys.} \textbf{58} 886

\bibitem{aut3} de Castro A S 2002 \textit{Phys. Lett. A} \textbf{305} 100

\bibitem{aut4} Nogami Y, Toyama F M and van Dijk W 2003 \textit{Am. J. Phys.} \textbf{71} 950

\bibitem{aut5} Gou Y and Sheng Z Q 2005 \textit{Phys. Lett. A} \textbf{338} 90

\bibitem{aut6} de Castro A S, Alberto P, Lisboa R and Malheiro M 2006 \textit{Phys. Rev. C} \textbf{73} 054309

\bibitem{aut7} Jia C-S, Guo P and Peng X-L 2006 \textit{J. Phys. A} \textbf{39} 7737

\bibitem{aut8} de Castro A S 2007 \textit{Int. J. Mod. Phys. A} \textbf{22} 2609

\bibitem{aut9} Castro L B, de Castro A S and Hott M 2007 \textit{Int. J. Mod. Phys. E} \textbf{16} 3002

\bibitem{aut10} Jia C-S, Guo P, Diao Y-F, Yi L-Z and Xie X-J 2007 \textit{Eur. Phys. J. A} \textbf{34} 41

\bibitem{aut11} Castro L B, de Castro A S and Hott M B 2007 \textit{EPL} \textbf{77} 20009

\bibitem{aut12} Qiang W-C, Zhou R-S and Gao Y 2007 \textit{J. Phys. A} \textbf{40} 1677

\bibitem{aut13} Bayrak O and Boztosun I 2007 \textit{J. Phys. A} \textbf{40} 11119

\bibitem{aut14} Soylu A, Bayrak O and Boztosun I 2007 \textit{J. Math. Phys.} \textbf{48} 082302

\bibitem{aut15} Soylu A, Bayrak O and Boztosun I 2008 \textit{J. Phys. A} \textbf{41} 065308

\bibitem{aut16} Zhang L-H, Li X-P and Jia C-S 2008 \textit{Phys. Lett. A} \textbf{372} 2201

\bibitem{aut17} Zhang F-L, Fu B and Chen J-L 2008 \textit{Phys. Rev. A} \textbf{78} 040101(R)

\bibitem{aut18} Akcay H 2009 \textit{Phys. Lett. A} \textbf{373} 616

\bibitem{aut19} Arda A, Sever R and Tezcan C 2009 \textit{Ann. Phys. (Berlin)} \textbf{18} 736

\bibitem{aut20} S.M. Ikhdair 2010 \textit{J. Math. Phys.} \textbf{51} 023525

\bibitem{aut21} Aydogdu O and Sever R 2010 \textit{Ann. Phys. (N.Y.)} \textbf{325} 373

\bibitem{aut22} Zarrinkamar S, Rajabi A A and Hassanabadi H 2010 \textit{Ann. Phys. (N.Y.)} \textbf{325} 2522

\bibitem{aut23} Oyewumi K J and Akoshile C O 2010 \textit{Eur. Phys. J. A} \textbf{45} 311

\bibitem{aut24} Hamzavi M, Rajabi A A and Hassanabadi H 2010 \textit{Phys. Lett. A} \textbf{374} 4303

\bibitem{aut25} Hamzavi M, Rajabi A A and Hassanabadi H 2010 \textit{Few-Body Syst.} \textbf{48} 171

\bibitem{aut26} Ikhdair S M and Sever R 2010 \textit{Appl. Math. Comput.} \textbf{216} 545

\bibitem{aut27} Ikhdair S M and Server R 2010 \textit{Appl. Math. Comput.} \textbf{216} 911

\bibitem{aut28} Aydogdu O and Sever R 2011 \textit{Phys. Lett. B} \textbf{703} 379

\bibitem{aut29} Candemir N 2012 \textit{Int. J. Mod. Phys. E} \textbf{21} 1250060

\bibitem{aut30} Zhang M-C and Huang-Fu G-Q 2012 \textit{Ann. Phys. (N.Y.)} \textbf{327} 841

\bibitem{aut31} Castro L B 2012 \textit{Phys. Rev. C} \textbf{86} 052201(R)

\bibitem{aut32} Hamzavi M and Ikhdair S M 2012 \textit{Can. J. Phys.} \textbf{90} 655

\bibitem{aut33} Agboola D 2012 \textit{J. Math. Phys.} \textbf{53} 052302

\bibitem{aut34} Hamzavi M, Eshghi M and Ikhdair S M 2012 \textit{J. Math. Phys.} \textbf{53} 082101

\bibitem{aut35} Ikhdair S M and Hamzavi M 2012 \textit{Few-Body Syst.} \textbf{53} 487

\bibitem{aut36} Guo J-Y 2012 \textit{Phys. Rev. C} \textbf{85} 021302

\bibitem{aut37} Chen S-W 2012 \textit{Phys. Rev. C} \textbf{85} 054312

\bibitem{aut38} Hamzavi M, Ikhdair S M and Ita B I 2012 \textit{Phys. Scr.} \textbf{85} 045009

\bibitem{aut39} Ikhdair S M and Server R 2012 \textit{Appl. Math. Comput.} \textbf{218} 10082

\bibitem{aut40} Akcay H and Server R 2013 \textit{Few-Body Syst.} \textbf{54} 1839

\bibitem{aut41} Thylwe K-E and Hamzavi M 2013 \textit{Phys. Scr.} \textbf{87} 025004

\bibitem{aut42} Ikhdair S M amd Falaye B J 2013 \textit{Phys. Scr.} \textbf{87} 035002

\bibitem{aut43} Liang H, Shen S, Zhao P and Meng J 2013 \textit{Phys. Rev. C} \textbf{87} 014334

\bibitem{aut44} Hamzavi M and Rajabi A A 2013 \textit{Eur. Phys. J. Plus} \textbf{128} 20

\bibitem{aut45} Hamzavi M and Rajabi A A 2013 \textit{Ann. Phys. (N.Y.)} \textbf{334} 316

\bibitem{aut46} Castilho W M and de Castro A S 2014 \textit{Ann. Phys. (N.Y.)} \textbf{340} 1

\end{thebibliography}
\end{document}